\documentclass[a4paper]{jpconf}

\usepackage{epsfig}
\usepackage{color}
\usepackage[utf8]{inputenc}
\usepackage[T1]{fontenc}
\usepackage{commath}

\begin{document}
\title{Relativistic hydrodynamics on graphics processing units}

\author{Jan~Sikorski$^1$, Sebastian~Cygert$^2$,
  Joanna~Porter-Sobieraj$^2$, Marcin~S{\l}odkowski$^1$,
  Piotr~Krzy\.{z}anowski$^1$, Natalia~Ksi\k{a}\.{z}ek$^1$ and Przemys{\l}aw~Duda$^1$}

\address{$^1$ Faculty of Physics, Warsaw University of Technology,
  Koszykowa 75, 00-662 Warsaw, PL}

\address{$^2$ Faculty of Mathematics and Information Science, Warsaw
  University of Technology, Koszykowa 75, 00-662 Warsaw, PL}

\begin{abstract}
  Hydrodynamics calculations have been successfully used in studies of
  the bulk properties of the Quark-Gluon Plasma, particularly of
  elliptic flow and shear viscosity. However, there are areas (for
  instance event-by-event simulations for flow fluctuations and
  higher-order flow harmonics studies) where further advancement is
  hampered by lack of efficient and precise 3+1D~program. This problem
  can be solved by using Graphics Processing Unit (GPU) computing,
  which offers unprecedented increase of the computing power compared
  to standard CPU simulations. In this work, we present an
  implementation of 3+1D ideal hydrodynamics simulations on the
  Graphics Processing Unit using Nvidia CUDA framework. MUSTA-FORCE
  (MUlti STAge, First ORder CEntral, with a~slope limiter and MUSCL
  reconstruction) and WENO (Weighted Essentially Non-Oscillating)
  schemes are employed in the simulations, delivering second
  (MUSTA-FORCE), fifth and seventh (WENO) order of accuracy. Third
  order Runge-Kutta scheme was used for integration in the time
  domain. Our implementation improves the performance by about
  2~orders of magnitude compared to a~single threaded program. The
  algorithm tests of 1+1D~shock tube and 3+1D~simulations with
  ellipsoidal and Hubble-like expansion are presented.
\end{abstract}

\vspace{-1cm}
\section{Numerical algorithms}
The equations of hydrodynamics can be written as a~system of
conservative equations
\begin{equation}\label{eqn:conservative}
  \pd{u}{t} + \sum\limits_{i} \pd{f_i(u)}{x_i} = 0
\end{equation}
where the sum runs over all spacial dimensions, and $u=u(x)$
represents a~conserved variable, which include the energy density,
momenta densities and conserved charge densities. The equation is
discretized on a~cartesian grid into an ordinary differential
equation
\begin{equation}\label{eqn:ode}
  \od{u}{t} = \sum\limits_i \frac{1}{\left| e_i \right|} \left(
    f_i\left(u\left(x-\frac{e_i}{2}\right)\right)
    - f_i\left(u\left(x+\frac{e_i}{2}\right)\right)
  \right) \equiv L(u)
\end{equation}
where $e_i$ are the primitive vectors, i.e.\ vectors between two
adjacent cells in the positive direction of dimension $i$. It is then
integrated using a~standard, third order Runge-Kutta algorithm. The
final discretized operator $L$ depends on a~number of neighbouring
cells on the lattice in all spacial directions.

% citeulike:2818714,
\subsection{MUSTA-FORCE scheme}
In order to obtain a general and accurate algorithm for estimating $L$
we use a~hybrid MUSTA (MUlti-STAge)
approach~\cite{Toro:2006:MMN:1163952.1163969, FLD:FLD980}. It utilizes
a~centered flux in a predictor-corrector loop, solving the intercell
Riemann problem numerically, without using a priori information about
waves that propagate in this system.

The algorithm is extended with the MUSCL scheme, which uses linear
interpolation inside the cells for second order accuracy. To reduce
oscillations that arise in the vicinity of strong gradients, we used
a~variety of slope limiting methods.

% 1059.65078,
% Xing:2006:HOW:1140818.1140826
\subsection{WENO scheme}
Due to high numerical cost and complexity of the MUSTA algorithm the
finite difference WENO scheme was also implemented~\cite{Shen2008,
  citeulike:6703557, Shu01highorder, Jiang95efficientimplementation,
  Qiu:2002:FDW:773502.942232}. It uses polynomial interpolations from
multiple stencils, a~convex combination of which, with suitably
non-linear coefficients, lead to a~scheme that stays high order in
smooth regions of the solution and discards stencils containing
shocks, thus avoiding spurious oscillations.

In our program the fifth (WENO5) and seventh (WENO7) order accurate
schemes were implemented.

\section{Technical aspects}
The code is written using the NVIDIA CUDA programming framework, which
produces binaries for NVIDIA graphics processing units. We compared
our implementation of the MUSTA-FORCE algorithm with an equivalent
implementation for the central processing unit. Using contemporary
hardware the GPU code executed approximately 200 times faster.

\section{Results}
We found that MUSTA-FORCE algorithm is more computationally expensive
and gives less accurate description of shock waves compared to WENO
algorithms. Thus only results obtained with WENO schemes are
presented.

\begin{figure}[htp]
\vspace*{-5pt}
$$\mbox{
\epsfig{width=8.cm,clip=,figure=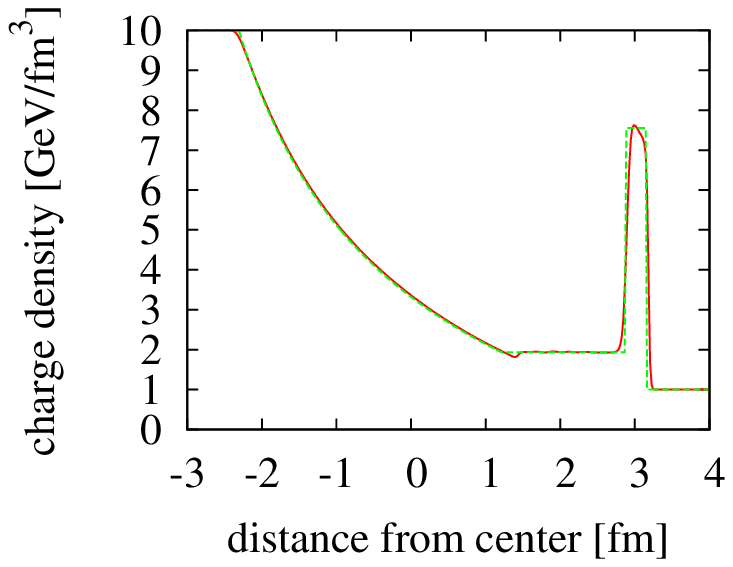}
\epsfig{width=8.cm,clip=,figure=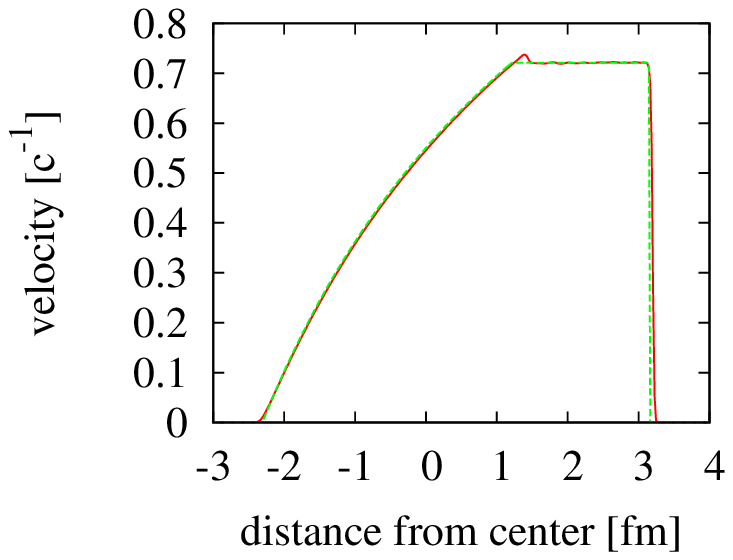}}$$
\vspace*{-20pt}
\caption{\label{fig:shock}Sod shock tube, charge density (left), velocity (right)} 
\end{figure}

Figure \ref{fig:shock} shows a~comparison of
the WENO5 scheme simulation with an analytical solution to the shock
tube problem. The results from WENO7 reproduce shock front better at
the cost of slight oscillations in the flat regions.

%%%%%%%%%%%%%%%%%%%%%%%%%%%%%%%%%%%%%%%%%%%%%%%%%%%%%%%%%%%%%%%%%%%%
%%%%%%%%%%%%%%%%%%%%%%%%%%%%%%%%%%%%%%%%%%%%%%%%%%%%%%%%%%%%%%%%%%%%

\begin{figure}[htp]
\vspace*{-5pt}
$$\mbox{
\epsfig{width=8.cm,clip=,figure=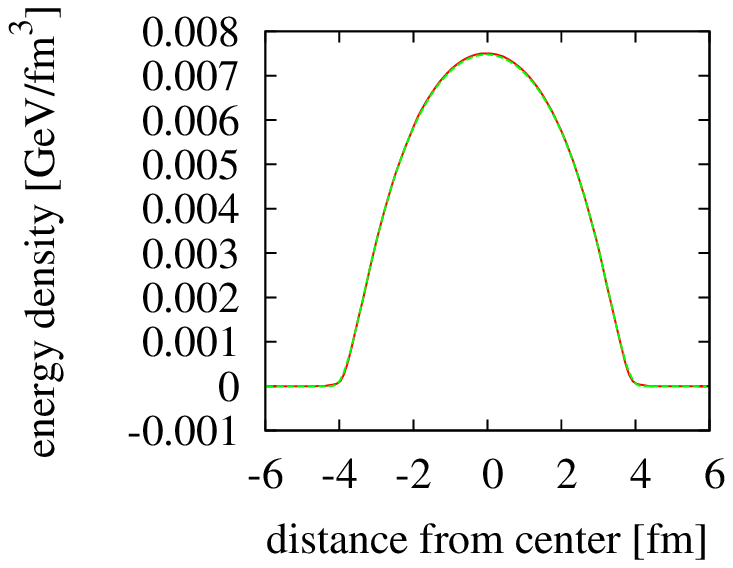}
\epsfig{width=8.cm,clip=,figure=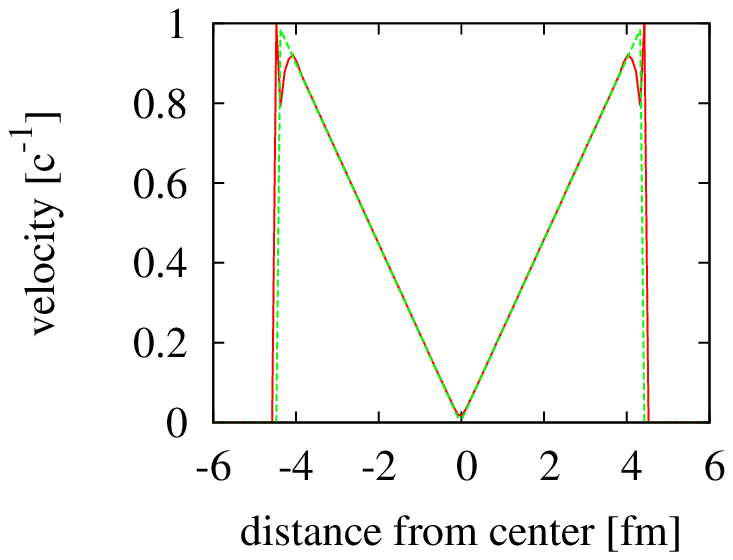}}$$
\vspace*{-20pt}
\caption{\label{fig:ellip}Ellipsoidal flow (WENO7), energy density (left), velocity (right).} 
\end{figure}

\begin{figure}[htp]
\vspace*{-5pt}
$$\mbox{
\epsfig{width=8.cm,clip=,figure=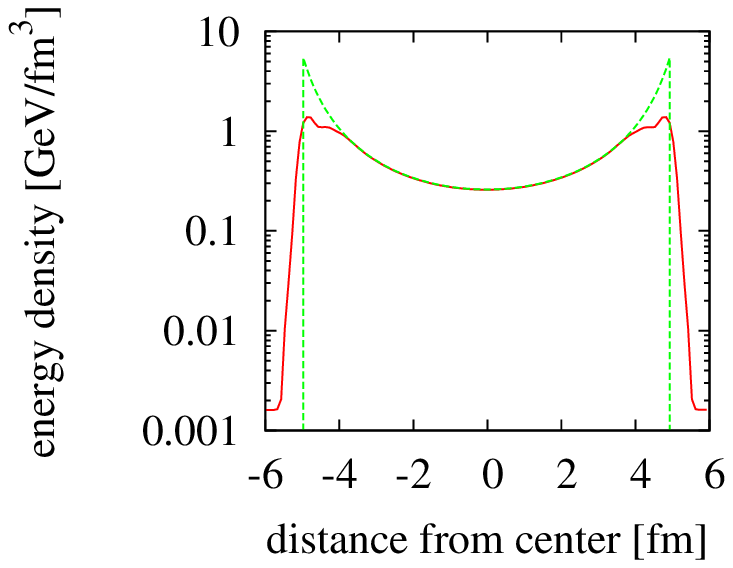}
\epsfig{width=8.cm,clip=,figure=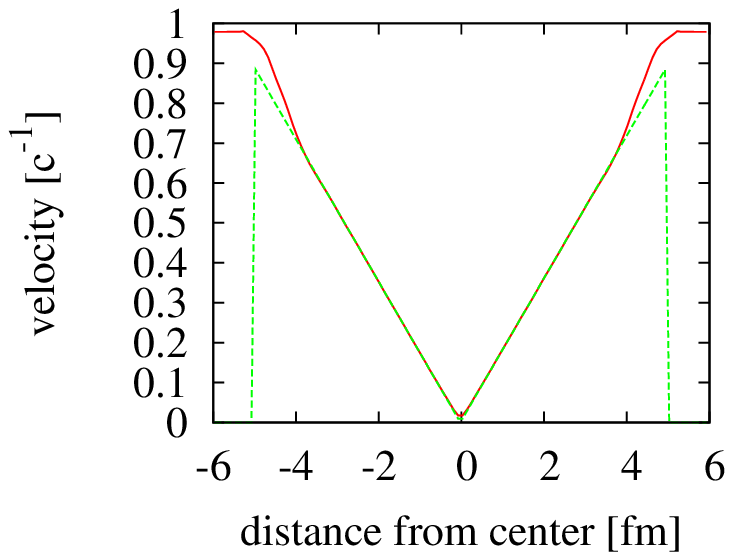}}$$
\vspace*{-20pt}
\caption{\label{fig:hubble}Hubble-like flow(WENO5), energy density (left), velocity (right).} 
\end{figure}

The following two simulations of ellipsoidal flow solution
\cite{Sinyukov:nucl-th0505041} and Hubble-like solution were done in
full 3+1D. On figures \ref{fig:ellip}--\ref{fig:hubble} one
dimensional sections are shown. Both WENO5 and WENO7 shown similiar
results in these tests---higher order simulations were slightly more
accurate in both cases.

%%%%%%%%%%%%%%%%%%%%%%%%%%%%%%%%%%%%%%%%%%%%%%%%%%%%%%%%%%%%%%%%%%%%
%%%%%%%%%%%%%%%%%%%%%%%%%%%%%%%%%%%%%%%%%%%%%%%%%%%%%%%%%%%%%%%%%%%%

\vspace{-0.5cm}

%%%%%%%%%%%%%%%%%%%%%%%%%%%%%%%%%%%%%%%%%%%%%%%%%%%%%%%%%%%%%%%%%%%%
%%%%%%%%%%%%%%%%%%%%%%%%%%%%%%%%%%%%%%%%%%%%%%%%%%%%%%%%%%%%%%%%%%%%

\end{document}